\documentclass[aip,rsi,amsmath,amssymb,preprint,
reprint
]{revtex4-1}

\usepackage{graphicx}% Include figure files
\usepackage{dcolumn}% Align table columns on decimal point
\usepackage{bm}% bold math
\usepackage[utf8]{inputenc}
\usepackage[T1]{fontenc}
\usepackage{mathptmx}
\usepackage{etoolbox}
\usepackage{siunitx}
\usepackage{xcolor}
\definecolor{myblue}{rgb}{0.12156862745098039, 0.4666666666666667, 0.7058823529411765}

\usepackage{hyperref}
\hypersetup{
    colorlinks=true,
    linkcolor=myblue,
    filecolor=magenta,      
    urlcolor=myblue,
    pdftitle={A self-referenced optical phase noise analyzer for quantum technologies},
    citecolor = myblue
}

\makeatletter
\def\@email#1#2{%
 \endgroup
 \patchcmd{\titleblock@produce}
  {\frontmatter@RRAPformat}
  {\frontmatter@RRAPformat{\produce@RRAP{*#1\href{mailto:#2}{#2}}}\frontmatter@RRAPformat}
  {}{}
}%
\makeatother
\begin{document}

% \preprint{}

\title{A self-referenced optical phase noise analyzer for quantum technologies}
% \title{A self-referenced optical phase noise analyzer}

\author{R. Freund}
\author{Ch. D. Marciniak}
\author{T. Monz}%
    \altaffiliation[Also at ]{Alpine Quantum Technologies GmbH, Innsbruck, Austria}
    \email{thomas.monz@uibk.ac.at}
\affiliation{ 
Universit{\"a}t Innsbruck, Institut f{\"u}r Experimentalphysik, Innsbruck, Austria
}

\date{\today}

\begin{abstract}
Second generation quantum technologies aim to outperform classical alternatives by utilizing engineered quantum systems. Maintaining the coherence required to enable any quantum advantage requires detailed knowledge and control over the noise the hosting system is subjected to. Characterizing noise processes via their power spectral density is routinely done throughout science and technology and can be a demanding task. Determining the phase noise power spectrum in leading quantum technology platforms, for example, can be either outside the reach of many phase noise analyzers, or be prohibitively expensive. In this work, we present and characterize a low-complexity, low-cost optical phase noise analyzer based on short-delay optical self-heterodyne measurements for quantum technology applications. Using this setup we compare two $\approx\SI{1}{\hertz}$ linewidth ultra-stable oscillators near \SI{729}{\nano\meter}. Their measurements are used as a baseline to determine and discuss the noise floor achieved in this measurement apparatus with a focus on limitations and their tradeoffs. The achieved noise floor in this all-stock-component implementation of an optical phase noise analyzer compares favourably with commercial offerings. This setup can be used in particular without a more stable reference or operational quantum system as sensor as would be the case for many component manufacturers.

\end{abstract}

\maketitle

\section{Introduction}
\label{sec:Intro}

Quantum technologies have made rapid progress from academic curiosity to a concentrated global effort at delivering devices that can outperform classical counterparts. This progression has seen the community expand to a broad base of backgrounds adopting and inventing techniques and technology from an equally wide range of disciplines. To achieve their desired task of outperforming classical alternatives, quantum technologies need to overcome challenges at the cutting edge of fabrication, engineering, and metrology.

All quantum technologies of the second generation aim to coherently control their constituent quantum systems and derive their advantage from this control. Here, we concentrate on quantum information processors but the same or similar statements are true in applications in e.g. quantum communication or metrology. The utility of quantum information processors as an example of second generation quantum technology hinges critically on their ability to produce and sustain coherence between their information carriers in the presence of a noisy environment and imperfect control. Precise knowledge of the noise that an information processor is subject to informs nearly all strategies to remove, reduce, or mitigate the detrimental effects noise has on operational fidelities or computational errors. Dephasing as an avenue to destroy coherence is an  ubiquitous error source in information processors that has in the past been primarily due to noise in e.g. the electromagnetic environment of the information carriers. However, state-of-the-art setups are now reaching environmental control at a level where instead the fluctuations in the control fields used to interact with information processors become limiting - a state that has been anticipated in the literature~\cite{ball2016role}. This expected scenario has now been explicitly identified as the main bottleneck in quantum enhanced metrology using ultra-cold atoms~\cite{pedrozo2020entanglement, cao2024multi,finkelstein2024universal}, as well as atomic quantum simulation, computation, and communication schemes within single-qubit or entangling operations~\cite{day2022limits,velazquez2024dynamical,marciniak2022optimal, nichol2022elementary}. Indeed, the physical dephasing noise an information processor is subject to can never be better than the intrinsic stability of the local oscillator (LO) it uses as a reference. Determining the phase noise floor of the local oscillator used to drive quantum operations is therefore now an important task. 

Noise spectroscopy using the quantum system as sensor has been studied in the context of quantum information processing~\cite{norris2018optimally,frey2020simultaneous} which has the benefit of probing the actual noise the processor is subject to. However, it runs the risk of conflating different types of detrimental influences, and critically requires an operational processor to begin with. While noise spectra find application in many protocols of quantum characterization, verification and validation~\cite{jiang2023sensitivity,nakav2023effect,day2022limits} or optimal control~\cite{green2013arbitrary,kang2023designing}, they are also an invaluable tool for component manufacturers that want to verify specifications without themselves operating a quantum information processor~\cite{di2010simple,day2022limits}.

The remainder of this manuscript is structured as follows: In section~\ref{sec:Background} we will outline the relevant background and context for determining the phase noise floor of an ultra-stable optical oscillator. In section~\ref{sec:Contribution} we will discuss the main contribution, and experimental setup and procedure of this work. In section~\ref{sec:Lasers} we compare two ultra-stable oscillators in use at the University of Innsbruck as a reference for the following section. We continue by outlining the noise floor of the device in section~\ref{sec:Limits} with a particular focus on the interconnected tradeoffs in system parameters and how they affect the floor. We conclude this manuscript in section~\ref{sec:Conclusion} by discussing system performance, how to improve it, and how this is positioned with respect to commercial offerings.

\section{Background}
\label{sec:Background}
Measurement of the phase (or frequency) noise of an oscillator via its noise power spectral density (PSD) is a routine task in science and engineering with commercial offerings available covering LO frequency ranges from static (DC) to many GHz. In these devices an internal reference is compared to the device under test and it is assumed that this reference contributes negligibly to the phase noise signal. However, once the stability of device and reference become comparable it is generally no longer possible to assign noise contributions to an individual device without relying on strong assumptions, for example that two devices are assumed to be identical. Another method is coherent cross-validation of more than two devices in for example a three-cornered-hat measurement~\cite{vernotte2016three}. A different approach with fewer assumptions in the situation where no suitably stable reference is easily available is to perform reference-free or self-referenced measurements. The oscillator under test is then compared with a delayed version of itself by means of a beat measurement in a homodyne or heterodyne detection scheme.

Either way, most phase noise analyzers operate entirely in the radio-frequency (RF) domain. However, local oscillators are frequently in the optical domain when the information carriers are based on atomic, molecular, or optical systems. In the optical domain no direct detection methods are available that can resolve the phase. Consequently, phase noise analyzers in that sense cannot be directly used. That said, typically phase noise varies on time scales much below the optical carrier frequencies themselves. In such instances the phase signal can be recovered by the phase-to-amplitude transduction in interferometers that converts the fluctuations around the optical frequency down to RF amplitude fluctuations amenable to standard detection schemes.

Complications arise when using interferometer-based optical phase detection due to the interferometer transfer function. This manifests as a periodic reduction of the phase signal amplitude with vanishing signal at multiples of the interferometer free spectral range (FSR). For medium stability oscillators this can be circumvented by using sufficiently large delay, effectively rendering the detection scheme incoherent. For ultra-low instability oscillators as used in quantum information processing this would require delay on a planetary length scale, and is consequently not feasible. Here we use the criterion on the maximum achievable delay to be much smaller than the coherence time of the oscillator as the boundary to 'ultra-stable'~\cite{richter1986linewidth}. For oscillators that are ultra-stable by that definition the coherence of the short delay interferometer plays a crucial role and has to be taken into account~\cite{ludvigsen1998laser}.

The complications for ultra-stable oscillators are further compounded by the observation that the measurement chain itself may contribute to the overall noise, or even dominate the signal. While careful selection and characterization of the measurement chain is important, there are detection schemes and post-processing techniques that can improve performance even below the bare noise characteristics of the phase noise analyzer. Prominently among these are differential detection which suppresses some amplitude noise coupling to the phase quadrature~\cite{wissel2022relative}, and utilizing the suppression of uncorrelated noise originating in the detectors via cross-correlation~\cite{walls1992cross}. Both of these come with a small overhead in system and post-processing complexity. 

A large body of work exists to deal with phase noise measurements of oscillators using short delay interferometers~\cite{salik2004dual,bai2021narrow} whose attractiveness for quantum technology application depend mainly on two factors: The achievable noise floor (performance) and availability (effort or financial). Many of the existing techniques do not reach sufficient sensitivity or rely on assumptions that are violated in ultra-stable oscillators as used in frequency metrology and quantum information processing~\cite{zhao2022narrow,bai2022lorentzian,huang2017precise,he2019high}, such as no structure aside from the transfer function. There do exist a number of methods that are suitable for these extremely low-noise oscillators under some conditions. For example, the (differential) detector of a short delay interferometer can be analyzed using a commercial RF phase noise analyzer. The downside to this is that it makes the measurement sensitive to the intrinsic noise of the interferometer photodetector. Another option is making use of the now available commercial optical phase noise analyzers. Both of these are viable options in terms of noise floor for many applications, but constitute a substantial financial investment. There are, however, a few alternative methods which impose less of a financial burden~\cite{li2019laser,gundavarapu2019sub,salik2004dual,XCOR} and can provide performance on par with commercial offerings, making phase noise measurements of ultra-low noise oscillators accessible to a wider community.

\section{Main contribution}
\label{sec:Contribution}

A key motivation of this work is to provide easy access to the construction of self-referenced phase noise analyzers for ultra-stable oscillators. Here we have adopted the correlated optical self-heterodyne method pioneered in the early 2000s~\cite{salik2004dual}, in particular its modern implementation (COSH)~\cite{XCOR}, to perform phase noise characterizations of the \SI{729}{\nano\meter} lasers that serve as the local oscillator for coherent operations on trapped $^{40}$Ca$^{+}$ optical qubit at the University of Innsbruck. In this scheme the demodulation of the heterodyne signal is performed digitally in post-processing, although analog demodulation is also possible. In our setup we have adapted the method to operation in near infrared rather than the telecom band, where fiber-based components are generally less available or of lower performance. In line with our objective to be instructive, we provide a thorough characterization of setup performance, limitations, and performance tradeoffs of operation parameters. We further discuss improvements to the implementation for different environments, and showcase the performance of the setup by comparing two \SI{1}{\hertz}-linewidth lasers operational at the University of Innsbruck, including cross-validation using a commercial phase noise analyzer and noise spectroscopy via spin locking on trapped-ion qubits. Our implementation of correlated self-heterodyne is cost-effective as it uses off-the-shelf stock components and hardware that is otherwise readily available in a typical laboratory environment.

\subsection{Experimental setup}
\label{sec:Setup}

\begin{figure*}[htb!]
    \includegraphics[width=\textwidth]{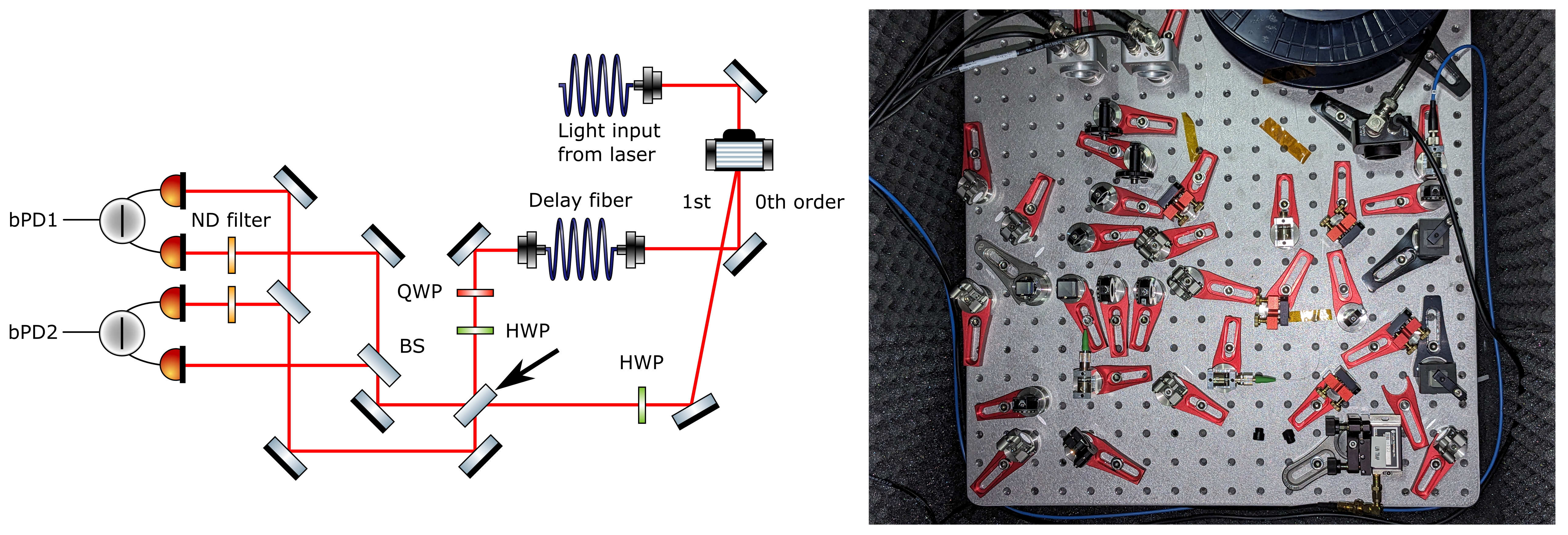}   
    \caption{Schematic of experimental setup. A phase-to-amplitude transduction is performed via a heterodyne Mach-Zehnder interferometer with an optical fiber delay line in one arm, and the frequency shift is generated using an acousto-optic modulator. HWP/QWP: Half/Quarter wave plate. BS: Beam splitter. ND: Neutral density. bPD: Balanced photodetector. The arrow points to the end of the interferometer after which noise is approximately common mode.} 
    \label{fig:Setup}
\end{figure*}

The phase noise analyzer consists of a free-space breadboard hosting the optical components shown in Fig.~\ref{fig:Setup} including the delay line, which constitutes the free-space analog of the setup shown in Ref.~\cite{XCOR}. The main sections are a frequency-shifted Mach-Zehnder interferometer with a delay line in one arm, followed by a path where two sets of balanced (differential) photodetectors (bPD) receive light from each output of the interferometer. The breadboard itself is situated inside an acoustic (vibration) shield made from \SI{16}{\milli\meter} three-layer fir plywood, \SI{2}{\milli\meter} lead foil, and  absorbing acoustic foam with a small penetration for in/egress of cables and optical fiber.

The frequency shift required for the heterodyne detection is provided by a Gooch \& Housego 3080-125 acousto-optic modulator (AOM) at a center frequency of \SI{80}{\mega\hertz}. The AOM is driven by either a Rigol DG4202 or DSG815 arbitrary waveform generator, which is then amplified. Adjusting the radio frequency power to the AOM allows to adjust the power splitting between the two interferometer arms to account for losses in the delay line. Polarization optics are used to compensate for mismatch in the electric field polarization due to the delay line and birefringence in the arms in order to superimpose the electric fields at the interferometer outputs in their degrees of freedom. Variable neutral density filters in front of the two bPDs are used to adjust the relative power of the input beams such that the average output voltage of the individual bPD coincides with its zero-input offset voltage. The optical power per interferometer arm is $\approx\SI{3}{\milli\watt}$ with $\approx$\SIrange{0.5}{2}{\milli\watt} of power in front of the bPDs. The delay line is given by a single-mode step-index optical fiber of either $\approx$\SI{12}{\meter} (S630-HP) or $\approx$\SI{1}{\kilo\meter} length (CS630-125-13/250).

The balanced photodetectors are MOGLabs PDA080, that is a customized version of PDA030 which are transimpedance amplified, operational amplifier-based RF circuits. They have a bandwidth of $\approx$\SI{80}{\mega\hertz}, and responsivities of $\approx$\SI{60}{\volt\per\watt} and $\approx$\SI{300}{\volt\per\watt} where we measure typically $\approx$\SIrange{0.3}{1}{\volt} signal amplitude limited by the optical power. The differential outputs are digitized simultaneously using two channels of a Rohde \& Schwarz RTM3004 oscilloscope. The data is post-processed and analyzed on a computer following mainly the \emph{pycosh} software package. For cross-validation one of the two differential outputs is fed into a Rohde \& Schwarz FSPN8 phase noise analyzer. The FSPN8 performs post-processing and analysis internally, but data can be exported for comparison.

Determining the noise floor of the device gives information about the smallest possible signal that could be measured and is fundamental for any measurement device. The noise floor is in part given by the measurement chain, while other parts can be dominated by environmental influences. There are two different approaches to determine the noise floor of the measurement chain itself: Either we characterize the individual parts of the chain separately, or we perform a measurement using a modified setup where the effect of the laser cancels on the detectors and the cumulative noise is measured. It is important to realize that the noise floor of the measurement chain depends both on the frequency difference between the two interferometer arms, and the signal amplitude. This is because the noise characteristics of each component are typically not spectrally flat, and because the ability to resolve small changes in phase is directly related to the signal to noise ratio (SNR) of signal (variable with optical input) to intrinsic noise (fixed).

We characterize the individual parts of the measurement chain in the first approach above. For the photodetectors we perform a dark measurement, with no incident light. We digitize a time trace of the detector output and add this to a perfect sine at the interferometer heterodyne frequency in post-processing. The amplitude of the added signal is set to a typical time trace with light. The analysis then proceeds analogous to a normal measurement. The RF drive of the AOM, which produces the frequency shift between the two interferometer arms, is characterized in a similar fashion. Finally, to characterize the cumulative effects of digitization noise and timing jitter in the oscilloscope we record time traces with both oscilloscope channels fed by a frequency reference (Rohde \& Schwarz SMA100B) whose characterized frequency noise is below our measurement chain floor. The resulting noise PSDs are then calculated as above.

The cumulative noise floor can be determined by performing a self-heterodyne beat measurement with equal delay on both arms. We performed these measurements with a free-space Mach-Zehnder interferometer bypassing the fiber as it is difficult to obtain the same delay using a fiber-based delay line. For identical interferometer arms the laser frequency noise does not contribute to the detected signal and the cumulative noise floor of the measurement chain remains, that is in particular excluding any additional noise contributed by the delay line.

In this work we have used both approaches with identical outcome. Therefore we present only the individual characterizations which directly inform improvements to the setup.

\section{Oscillator comparison}
\label{sec:Lasers}

We record and compare the frequency power spectral density of two ultra-stable optical oscillators near \SI{729}{\nano\meter} that are used to drive coherent qubit operations on an electric quadrupole transition in a trapped-ion quantum computer. The two oscillators are based on a frequency-converted Titanium-Sapphire laser (TiSa) and a stabilized diode laser (DL), respectively. Both lasers are locked to high-finesse cavities of finesse $\approx$200.000 and feature several servo loops that stabilize the output frequency such that they have linewidths of $\approx$\SI{1}{\hertz} as determined via direct beat measurements and a three-cornered-hat measurement with a third DL-type laser of identical make. However, the details of the servo system as well as their makeup are different, such that the spectral content is expected to be distributed differently between the two. The frequency noise power spectral densities presented here for a given configuration and set of parameters will be used as reference for illustrating how the performance of the phase noise analyzer is affected by changing its setting.

\begin{figure*}[hbt!]
    \centering
    \includegraphics{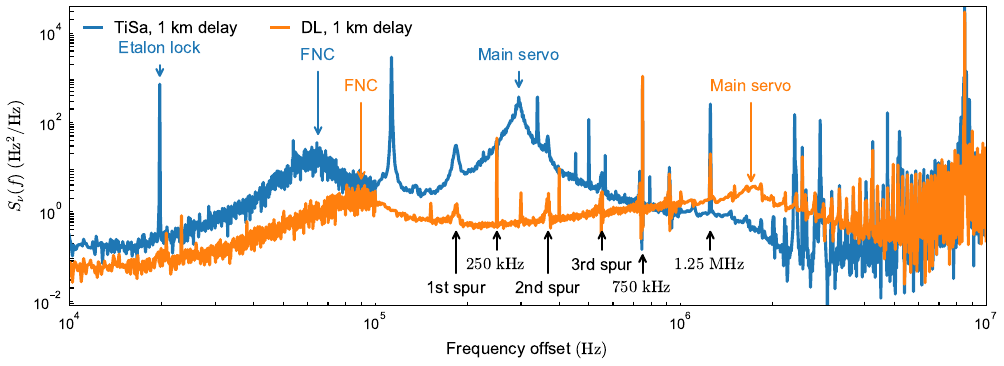}
    \caption{Single-sided frequency noise power spectral density $S_{\nu}(f)$ of two ultra-stable oscillators near \SI{729}{\nano\meter} as measured using the presented setup. The delay line was a \SI{1}{\kilo\meter} single-mode fiber producing an interferometer free spectral range of $\approx$\SI{184}{\kilo\hertz}. Spikes at this Fourier frequency and its multiples are artifacts of data evaluation and can be removed if necessary (see main text). The first three are denoted as 'spur' and occur in both traces. RF pickup, marked at their respective frequencies in black, also feature in both traces. The frequency range displayed is chosen to roughly correspond to the region wherein the laser frequency noise dominates the detected signal.}
    \label{fig:Comparison}
\end{figure*}

Within the spectral region displayed in Fig.~\ref{fig:Comparison} the frequency noise PSDs for both lasers show similar behaviour. Both lasers feature a broad servo bump at $\approx$\SI{65}{\kilo\hertz} and $\approx$\SI{90}{\kilo\hertz} for TiSa and DL, respectively, originating from the fiber noise cancellation (FNC) feedback loop~\cite{foreman2007remote} that cancels additional phase noise imprinted on the laser through the fiber that connects the oscillator source with the optical table. This peak will be present on the light as seen by the qubits and as such is genuine noise for our analysis. Both lasers also feature a second servo bump from the main frequency stabilization sitting at $\approx$\SI{300}{\kilo\hertz} and $\approx$\SI{1.8}{\mega\hertz} for the TiSa and DL, respectively. Both lasers' PSDs fall at higher frequencies before reaching the $\approx f^2$ high-frequency noise floor. The TiSa spectrum features an additional narrow but high noise spike at \SI{19.8}{\kilo\hertz} from an etalon pre-lock.

There are understood measurement artifacts in both traces in addition to the structure originating from the laser. Firstly, there are a number of peaks attributed to RF pickup in the measurement chain. These can be identified from dark measurements at \SI{250}{\kilo\hertz}, \SI{750}{\kilo\hertz} and \SI{1.25}{\mega\hertz}. These RF pickups can only be removed under assumptions or by eliminating them through stronger shielding or removal of the source. Secondly, both spectra feature periodic spurs at multiples of the interferometer FSR of $\approx$\SI{184}{\kilo\hertz} which become increasingly compressed on the log-log scale display. At multiples of the FSR the interferometer signal vanishes in the absence of detector noise for infinite measurement bandwidth. For finite bandwidth it is suppressed but remains finite. This modulation is removed in post-processing by multiplying with a processing gain following Eq.~(22) in Ref.~\cite{XCOR}. Assuming a perfect detector leads to noise spurs when the detector noise is above the (finite-bandwidth) estimate and the spectrum overshoots after compensation. The overshoot grows and then saturates towards higher frequencies as the noise floor washes out the interferometer transfer function minima. Note that dark measurement traces will always show large noise spurs at multiples of the FSR. For compatibility, all dark measurements are also multiplied with the processing gain and since those measurements do not have the periodic modulation from the interferometer, they will always overshoot maximally.

The position of these spurs allows determination of the delay via the interferometer FSR. Spurs are measurement artifacts that may be undesirable. It is worth pointing out that they can be removed for example if two different delays are available. In that case, their position will be different between the two measurements and the underlying value of the PSD in the contaminated regions can be found from the second spectrum, assuming the different delay does not raise the measurement floor above the signal value.

\begin{figure*}[htb!]
    \centering
    \includegraphics{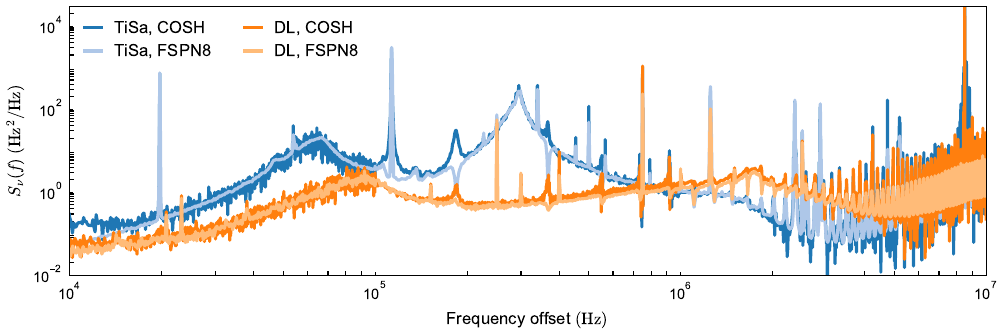}
    \caption{Spectra for the TiSa (blue) and DL (orange) lasers produced using correlated self-heterodyne and a commercial phase noise analyzer FSPN8 (pale blue and pale orange) for cross-validation. The measurement record of the FSPN8 is substantially longer, which reduces variance at low frequencies. At the optical power levels used here the uncorrelated optical detector (bPD) noise floor limits the spectrum on the FSPN8 roughly at the same level as the oscilloscope noise floor using correlated self-heterodyne.}
    \label{fig:PNA}
\end{figure*}

The frequency noise PSDs produced by the setup were cross-validated using a commercial phase noise analyzer (FSPN8) whose input signal is the balanced output of one of the two photodetector pairs. The data of the commercial device shown in Fig.~\ref{fig:PNA} is compensated for the modulation by the FSR and cross-correlated only for the noise induced by the FSPN8 itself. The measurement noise originating at the balanced photodetector can not be reduced in this fashion.

\section{Measurement limitations and tradeoffs}
\label{sec:Limits}

What limits the noise floor of the phase noise analyzer is determined by a number of factors which depend on both the environment and available equipment. In this section we will discuss how changes in the system parameters qualitatively affect the resulting noise floor and the tradeoffs between them, grouping them by parameters of the (post-)processing and the physical apparatus. A summary of operational parameters used in the measurement chain of this work is shown in Tab.~\ref{tab: parameters}.

\begin{table}

\begin{center}
    \begin{tabular}{c|c}
    Quantity & Value\\\hline
        Sampling rate & \SI{312.5}{\mega \hertz} \\
        Samples per trace & $40\cdot10^6$\\
        Digitization bit depth (specified) & 10 bit \\
        Optical power per branch & $~\SI{3}{\milli \watt}$ \\
        Optical power before PD & \SIrange{0.5}{2}{\milli \watt} \\
        Voltage amplitude on bPD & \SIrange{0.3}{1}{\volt}\\
        noise equivalent power of PD & \SIrange{170}{330}{\pico \watt / \sqrt{\hertz}}\\
        RIN to phase noise conversion factor & $3 \cdot 10^{-1} \, \text{rad}^2$ \\
        Dynamic range & \SIrange{0.01}{10}{\mega \hertz}
    \end{tabular}
    
    \vspace{1em}
    
    \begin{tabular}{c|c|c}
        Frequency range & RBW & Number of \\
         & & cross corr. \\
        \hline
        \SIrange{1}{10}{\kilo \hertz} & \SI{10}{\hertz} & 1\\
        \SIrange{10}{100}{\kilo \hertz} & \SI{100}{\hertz} & 12\\
        \SIrange{0.1}{1}{\mega \hertz} & \SI{1}{\kilo \hertz} & 122\\
        \SIrange{1}{10}{\mega \hertz} & \SI{10}{\kilo \hertz} & 1228\\
    \end{tabular}
\end{center}
\caption{Measurement chain operational parameters used for the data taken in this manuscript.}
\label{tab: parameters}
\end{table}

The frequency range which can be considered using correlated self-heterodyne is fundamentally limited towards higher Fourier frequencies by the interferometer heterodyne frequency. The offset frequency shifts the detected frequency noise away from the typically strong noise close to DC. However, as the offset from the carrier becomes appreciable this noise may again contaminate the phase noise measurement. This contamination breaks base assumptions in the derivation of the phase signal - critically it induces asymmetry with respect to the carrier. Larger offset frequencies can be faithfully measured by increasing the heterodyne separation at the cost of higher bandwidth and memory needed in the detection chain. There is no such fundamental limit towards zero offset frequency. However, practically only a finite amount of data is available which limits the bandwidth of the phase noise analyzer as well as raising the (finite-bandwidth) noise floor due to fewer averages.

\subsection{(Post-)processing parameters}
\label{sec:ProcessingParameters}

Data recorded in the measurement chain consists of pairs of time traces of voltages from the balanced photodetectors with corresponding sampling rate, sample length and digitization resolution used here stated in Tab. \ref{tab: parameters}. These time traces are post-processed using the \emph{pycosh} software package. The phase of the trace is extracted by a discrete Hilbert transformation followed by removing 2\% of the record at the beginning and end of the transformed signal to avoid distortions at the endpoints. A Fourier transformation, cross-correlation, and potentially averaging gives the final spectrum. Any record of data will be of a finite size, limited either by acquisition system memory or recording time. There are tradeoffs in how this finite record is used in the post-processing during the various steps. While using all available data in a single Fourier transform allows to probe the PSDs at lower frequencies, the produced estimator of the underlying spectrum is generally poor. Consequently, typically the record is broken into smaller sections which are each transformed and then averaged over. This improves the estimator statistics at the cost of the low frequency end. Likewise, high sampling rates lead to more data generated and larger spectral bandwidth, but reduce the record temporal extend at fixed memory, as well as causing additional overhead in processing. For a fixed bandwidth and resolution, cross-correlation and potential averaging increase the amount of measurement time and memory required.

As part of the suppression of detector chain noise we perform cross-correlation of Fourier coefficients of the different detectors. We point out that while originally intended to reduce the influence of noise from the detectors, cross-correlation will also suppresses contributions from any other uncorrelated noise in the measurement chain that is common to both interferometer arms. A relevant example of this is digitization noise or jitter between the channels of the oscilloscope if they are uncorrelated. More cross-correlation can suppress detection chain noise further but it reduces the available measurement bandwidth. In this work we only cross-correlate instead of averaging since for suppressed uncorrelated noise the two operations are identical. High frequency noise can be readily identified using data stretching over a short temporal extent. Therefore, \emph{pycosh} uses adaptive bandwidths for the two competing requirements, cross-correlating more frequently at higher Fourier frequencies. As a result the variance of the produced PSDs increases discontinuously at the transition between bandwidth windows. The (non-generic) number of cross-correlations and the resolution bandwidth for the corresponding frequency range used in this work are shown in Tab. \ref{tab: parameters}.

\begin{figure*}
    \centering
    \includegraphics{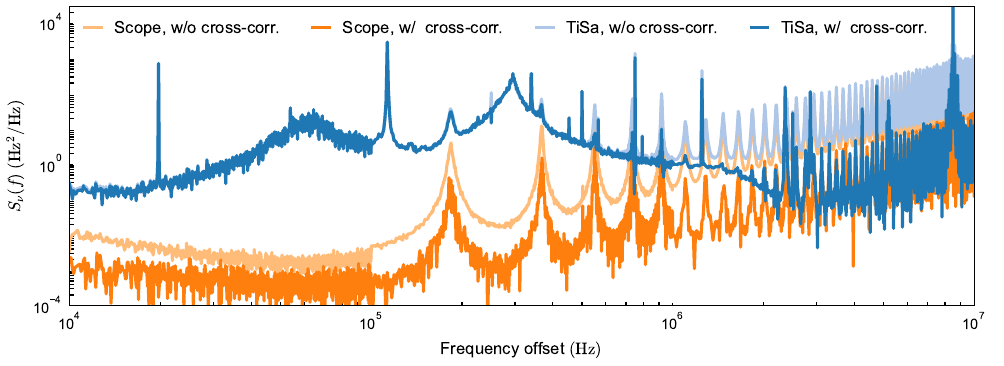}
    \caption{Reconstructed frequency noise PSDs for measurements with and without light showcasing the effect of cross-correlation. Dark measurements of the effect of the oscilloscope using a high-quality frequency reference show a suppression of the overall noise for the cross-correlated two channel measurement  (orange tones). The suppression increases towards higher frequencies. Peak structures arise from interferometer transfer function removal. The optical frequency noise PSD for a cross-correlated signal falls below the output of a single balanced photodetector (blue tones) for high frequencies.}
    \label{fig:CrossCorrelation}
\end{figure*}

When and how much cross-correlation benefits the overall signal depends on the relative size of the other noise contributions. In Fig.~\ref{fig:CrossCorrelation} we illustrate two cases with and without light. When performing a dark characterization measurement of the oscilloscope using a high-quality frequency reference we see that cross-correlation improves the noise floor for all considered frequencies with a stronger effect towards higher offsets. The frequency noise PSD of the TiSa laser does not benefit from cross-correlation at lower frequencies as the laser noise itself dominates over the detection chain contribution. However, at higher offset frequencies the laser noise without cross-correlation is dominated by the oscilloscope noise and the cross-correlated laser noise spectrum falls below the oscilloscope noise floor without cross-correlation.

\subsection{Apparatus parameters}
\label{sec:ApparatusParameters}

The noise floor of the phase noise analyzer depends strongly on the signal to noise ratio of the recorded time traces. The photodetectors' dark noise sets a lower bound on the optical power that is required to obtain a useful signal. The power required to overcome the dark noise for unit SNR is frequently quoted as noise equivalent power (NEP) or can be calculated from the voltage fluctuations without input light given the detector responsivity (Tab. \ref{tab: parameters}). Increasing the optical power level reduces the contribution of the photodetector noise to the overall device noise floor by increasing the SNR. A different way to increase the noise sensitivity of the setup at fixed input power, that is fixed photodetector noise floor, is via the interferometer delay.

The interferometer's phase-to-amplitude transduction, that is how much amplitude change per phase increment, is determined by the overall delay. The transducer slope increases with delay length, such that a short delay will produce a smaller amplitude change for a fixed phase noise strength and Fourier frequency compared to a longer delay. While longer delay makes weak noise more discernible, it lowers the interferometer FSR and thus increases the number of (possible) noise spurs in a given bandwidth. Additionally, longer delay lines suffer from enhanced acoustic noise pickup which influences the frequency noise floor at acoustic frequencies. This can in principle be mitigated by enhanced acoustic shielding or advanced mounting solutions~\cite{jeon2023palm}, but presents an important design consideration if the spectrum at low offset frequencies is desired. Fiber noise cancellation cannot be used to reduce the effect of acoustics (vibration), however, since the feedback loop does not distinguish between the desired phase signal and added acoustic noise. Should acoustic (vibration) shielding be sufficient for long delay lines then thermodynamic noise presents the next limiting factor~\cite{bartolo2012thermal, bilenko2004mechanical}.

Relative amplitude noise (RIN) cannot be distinguished from the phase signal after the interferometer transduction. While the balanced detection scheme of correlated self-heterodyne reduces the sensitivity to some types of RIN~\cite{wissel2022relative} ('1f noise', that is noise close to the heterodyne frequency), the suppression is finite only and depends on the device characteristics~\cite{XCOR}. Differential detection increases sensitivity to other types of RIN ('2f noise', that is noise close to twice the heterodyne frequency) non-linearly~\cite{lessing2013suppression,wissel2022relative}. However, this enhancement is only relevant if there are significant intensity noise sources at high frequencies, for example originating from a laser medium resonance or amplitude-to-phase transduction in electro-optic modulators. The spectral frequency noise contributions from shot noise-limited RIN fall with the average optical power incident on the photodetectors~\cite{day2022limits}. The equivalent shot noise of the light is typically orders of magnitude below the RIN contribution still and can therefore be safely neglected given typical power levels required for good detector signal to noise ratios.

Digitization noise and timing jitter in the recording equipment have a similar effect to photodetector dark noise. When two nominally different detector voltages fall within the same bin of the analog-to-digital converter they are mapped to the same value and will therefore not be resolved. Digitization noise can be reduced by using the full dynamic range of the digitizer, which requires the signal amplitude to be sufficiently stable. Increasing the bit depth of the digitizer decreases the impact of this effect on the device noise floor. Many digitizers offer 'enhanced bit depth' or 'high resolution' modes. These sacrifice temporal resolution, therefore bandwidth and number of samples, for higher effective voltage resolution. Timing jitter creates apparent phase jumps by misidentifying the temporal position of measurement points, therefore directly producing excess phase noise.

Any excess noise introduced through the element creating the frequency offset between the two interferometer arms will be indistinguishable from the measurement signal. As such, amplitude-to-phase coupling in the transducer AOM will add directly to RIN and phase noise from the RF source, amplifier or AOM will directly imprint on the PSD. Since these do not scale with signal amplitude the only way to reduce them is by controlling the noise in the devices themselves.

\section{Discussion \& conclusion}
\label{sec:Conclusion}

\begin{figure*}[htb!]
    \centering
    \includegraphics{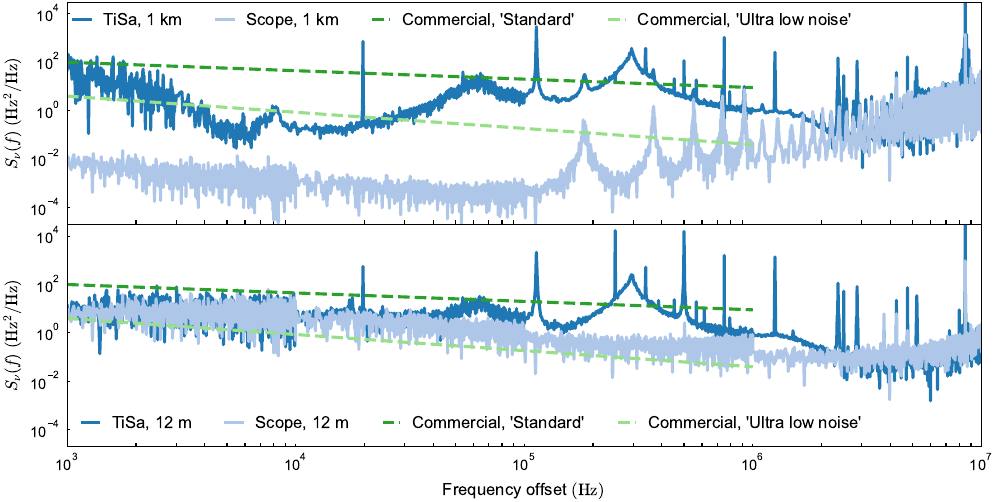}
    \caption{Comparison of noise floors (pale blue) and reconstructed frequency noise PSDs (dark blue) for TiSa laser for a $\approx$\SI{1}{\kilo\meter} (top) and $\approx$\SI{12}{\meter} (bottom) delay line. Below $\approx$\SI{5}{\kilo\hertz} excess phase noise ($\approx f^{-2}$) from acoustic (vibration) coupling sets the noise floor for the longer delay line (see Fig.~\ref{fig:SpinLocking}). Both are limited above $\approx$\SI{3}{\mega\hertz} by the oscilloscope noise floor. The shorter delay line is limited by the noise floor below $\approx$\SI{40}{\kilo\hertz}. For comparison specifications for commercial optical phase noise analyzer~\cite{OEwavesOE4000} are indicated (green).}
    \label{fig:NoiseFloors}
\end{figure*}

The cumulative noise floor of the setup for two different delay lengths is shown in Fig.~\ref{fig:NoiseFloors}. We used an optical power of $\approx$\SI{1}{\milli\watt} in front of each PD at which point phase noise contributions from our detectors, AOM and RF chain, and RIN were negligible when using cross-correlation.

\begin{figure}[b]
    \centering
    \includegraphics{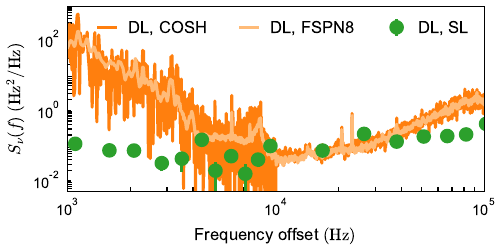}
    \caption{Comparison of phase noise analyzer data and noise spectroscopy via spin locking (SL) on trapped-ion qubit driven by the same oscillator at acoustic (vibration) noise floor cross-over. Noise estimates using trapped-ion qubits as sensors are consistent with the phase noise analyzer above the crossover, then fall below the phase noise analyzer estimates in the region where acoustic frequencies are suspected to dominate the phase noise analyzer floor.}
    \label{fig:SpinLocking}
\end{figure}

For the short delay line of $\approx$\SI{12}{\meter} the noise floor in the considered region is dominated entirely by the oscilloscope. For the longer delay line of $\approx$\SI{1}{\kilo\meter} there is excess phase noise below $\approx$\SI{5}{\kilo\hertz} following $\approx f^{-2}$. We attribute this to acoustic (vibration) coupling to the long fiber spool. Using the FSPN8 we have probed the frequency noise PSD down to \SI{10}{\hertz} and found good qualitative agreement in the structure of the PSD with vibration measurements performed in the same laboratory several years prior~\cite{pogorelov2021compact}. Further evidence of this noise floor stemming from the setup rather than being part of the oscillator's PSD is provided by performing noise spectroscopy via spin locking~\cite{Kranzl} based on the original magnetic resonance techniques~\cite{PhysRev.135.A1099,Slichter:801180,bodey2019optical}. Measurements using this technique on a trapped-ion quantum computer~\cite{pogorelov2021compact} show an estimated noise PSD in agreement with the phase noise analyzers, but drop below the PSD of the optical phase noise analyzer below $\approx\SI{5}{\kilo\hertz}$ as seen in Fig.~\ref{fig:SpinLocking}. We further see direct coupling in this spectral region by applying acoustic perturbations outside, suggesting that the approximately two orders of magnitude suppression of acoustic (vibration) coupling through the shield are insufficient at long delays. As expected, the spin locking estimates also miss the FNC servo at roughly \SI{90}{\kilo \hertz} since the light going to the experiment is split off before the FNC. The characterized noise floor for the long delay line is below the specifications of a commercial optical phase noise analyzer~\cite{OEwavesOE4000}in the region between \SI{10}{\kilo\hertz} and \SI{1}{\mega\hertz} for both standard and ultra low noise floor. In Tab. \ref{tab: parameters}, this range is referred to as dynamic range, which would grow larger for less stable oscillators under test.

We have previously mentioned that RIN is negligible for our measurements. However, a more careful discussion of RIN contributions is warranted since it is a source of easy confusion. A phenomenological RIN-to-frequency noise conversion $S_{\nu,\text{RIN}}(f) = f^2 \alpha \text{RIN}(f)$ with a conversion factor $\alpha$~\cite{XCOR} was determined at \SI{15}{\kilo\hertz} and \SI{50}{\kilo\hertz} offset frequency, chosen as a part of the spectrum where the measured noise level is low and with little structure, to be $\alpha~\approx 10^{-1}\, \text{rad}^2$, also shown in Tab. \ref{tab: parameters}. In Ref.~\cite{XCOR} a value of $\approx 10^{-4}\, \text{rad}^2$ at $\approx\SI{1}{\mega\hertz}$ was found. Following the exposé in Ref.~\cite{wissel2022relative} (see in particular Eq.~(39a)) and assuming perfectly identical, ideal detectors placed directly behind the beam combiner, estimated power ratios, contrast and reflection/transmission coefficients for our components, as well as solely correlated $1f$ RIN we find a theory estimate for $\alpha\approx 0.08$. This is surprisingly close to our measured value considering the double-balanced detection scheme with more (imperfect) beamsplitters, imperfect differential photodetectors, and multiplicative $2f$ RIN. The main contributors to $\alpha$ are low interferometric contrast caused by mode matching, which is enforced to be high in fiber-based detectors, as well as the quality of the beamsplitter(s) used.

However, it is important to point out that the size discrepancy in $\alpha$ between fiber-based and free-space implementation was not limiting the measurements. We separately determine the RIN present for the oscillators and converted it to frequency noise PSDs using our measured $\alpha$. The RIN-equivalent phase noise PSD is lower than 3\% of the laser phase contributions and therefore presents no limitation to the data shown despite the comparatively large coupling. This highlights that while the coupling could be reduced with better components and more careful mode matching, such improvements would only become necessary for substantially better oscillators.

We note that RIN coupling into the phase noise may also present a noise floor that is not shown here since it depends also on the spectral makeup of the oscillator, in particular for $2f$ RIN, rather than solely the device. Indeed, for some settings the RIN coupling floor intersects the oscilloscope noise floor at some frequencies.

As has been demonstrated, the cumulative noise floor of the phase noise analyzer can be reshaped by adjusting the many interconnected parameters. In our setup the most sensible revision would be a lower noise digitization solution either using a more advanced oscilloscope or a dedicated data acquisition card. We anticipate that this is going to be common for applications where sufficient optical power for a measurement can be supplied, since high quality digitizers are more expensive than high quality detectors in the optical domain.

Another option would be to tackle acoustic (vibration) coupling given that the low frequency side of the spectrum is of greater interest. The most straight-forward way to do this is to move to a heavier, sandwiched shield construction. However, this would mean the setup becoming more difficult to use, and laboratory space is often restricted. While the current revision of the setup consists of discrete components, it could be substantially smaller in footprint by employing a dedicated, monolithic design. Such designs are relatively inexpensive and would increase the amount of acoustic (vibration) shielding possible at equal total weight or footprint. However, the main limitation is the delay line, which is difficult to reduce in size for long delays. Overall, the delay line should be chosen as short as possible considering the other limitations to reduce phase noise introduced by the fiber.

In this work we have described and characterized an optical phase noise analyzer operating near \SI{729}{\nano\meter}. The setup works with inexpensive, conventional stock components and can be operated at any optical wavelength for which components are available. Despite the inexpensive construction, we see performance comparing very favourably with commercial phase noise analyzers. We expect that this development will make phase noise PSDs of ultra-stable reference oscillators available to the wider community, in particular the quantum industry. This can range from experimentalists wanting to characterize or debug their systems, to component manufacturers without direct access to a quantum information processor.

\section*{Acknowledgements}

The authors would like to acknowledge many helpful discussions with Manfred Hager, TOPTICA Photonics AG, as well as Robert Scholten, MOG Laboratories Pty Ltd, and Zhiquan Yuan, Caltech. We are grateful to Rohde \& Schwarz GmbH \& Co. KG for generously providing the FSPN8 phase noise analyzer for use in this publication. The authors thank Michael Meth and Lukas Postler for maintaining the oscillators characterized in this study.

The authors gratefully acknowledge funding by the U.S. ARO Grant No. W911NF-21-1-0007. We also acknowledge funding by the Austrian Science Fund (FWF),
through the SFB BeyondC (FWF Project No. F7109), 
by the Austrian Research Promotion Agency (FFG) contract 897481 and 896039 (Next Generation EU),
and has received funding from the European Union’s Horizon Europe research and innovation programme under grant agreement No 101114305 (“MILLENION-SGA1” EU Project)
and the IQI GmbH. 
Views and opinions expressed are, however, those of the author(s) only and do not necessarily reflect those of the European Union or the European Commission. Neither the European Union nor the granting authority can be held responsible for them.

\section*{Conflicts of interest}

The authors have no conflicts to disclose.

\section*{Author contributions}
\textbf{Robert Freund:} Investigation (lead); Writing (supporting); Formal Analysis (equal)
\textbf{Christian D. Marciniak:} Investigation (equal); Supervision (equal); Writing (lead); Formal Analysis (equal)
\textbf{Thomas Monz:} Funding Acquisition (lead); Supervision (equal); Writing/Review (supporting)

\section*{Data Availability Statement}

The data that support the findings of this study are openly available in Zenodo at \url{https://doi.org/10.5281/zenodo.8434568}, reference number 10.5281/zenodo.8434568.

% \appendix

% \section{Appendixes}

\bibliographystyle{apsrev4-1}
\bibliography{xcor}% Produces the bibliography via BibTeX.

\end{document}